\documentclass[letterpaper,amsmath,
  amssymb,aps,prd,nofootinbib,
twocolumn,
a4paper,
superscriptaddress,altaffilsymbol]{revtex4-1}

\usepackage{amsmath}
\usepackage{amsfonts}
\usepackage{amssymb}
\usepackage{graphicx}
\usepackage{psfrag}
\newcommand{\be}{\begin{equation}}
\newcommand{\ee}{\end{equation}}
\newcommand{\ba}{\begin{eqnarray}}
\newcommand{\ea}{\end{eqnarray}}
\newcommand{\beq}{\begin{equation}}
\newcommand{\eeq}{\end{equation}}
\newcommand{\beqa}{\begin{eqnarray}}
\newcommand{\eeqa}{\end{eqnarray}}

\newcommand{\eref}[1]{Eq.~\eqref{#1}}

\newcommand*{\cref}[1]{Chapter~\ref{#1}}

\renewcommand{\prd}{Phys. Rev. D}
\renewcommand{\apj}{ApJ}

\newcommand{\jcap}{JCAP}

\begin{document}

%\begin{flushright} {\footnotesize IC/2007/001
%\\ HUTP-07/A0002}  \end{flushright} 

\title{Einstein's signature in cosmological large-scale structure}
\author{Marco Bruni}
\email{marco.bruni@port.ac.uk}
\affiliation{Institute of Cosmology and Gravitation, University of
  Portsmouth, Dennis Sciama Building, Portsmouth PO1 3FX, United Kingdom.}
\author{Juan Carlos Hidalgo}
\email{hidalgo@fis.unam.mx}
\affiliation{Instituto de Ciencias F\'{\i}sicas, Universidad Nacional
Aut\'onoma de M\'exico, Cuernavaca, Morelos, 62210, Mexico}
\affiliation{Institute of Cosmology and Gravitation, University of
  Portsmouth, Dennis Sciama Building, Portsmouth PO1 3FX, United Kingdom.}
\author{David Wands}
\email{david.wands@port.ac.uk}
\affiliation{Institute of Cosmology and Gravitation, University of
  Portsmouth, Dennis Sciama Building, Portsmouth PO1 3FX, United Kingdom.}

\date{\today}
\begin{abstract}

We show how the non-linearity of general relativity generates a characteristic non-Gaussian signal in cosmological large-scale structure that we calculate at all perturbative orders in a large scale limit.
Newtonian gravity and general relativity provide complementary theoretical frameworks for modelling large-scale
structure in $\Lambda$CDM cosmology;
a relativistic approach is essential to determine initial conditions which can then be used in Newtonian simulations studying the non-linear evolution of the matter density.
Most inflationary models in the very early universe predict an almost Gaussian distribution for the primordial metric perturbation, $\zeta$.
However, we argue that it is the Ricci curvature of
comoving-orthogonal spatial hypersurfaces, $R$, that drives structure
formation at large scales.  
We show how the non-linear relation between the spatial curvature, $R$, and the 
metric perturbation, $\zeta$, translates into 
a specific non-Gaussian contribution to the initial comoving matter density that we calculate for the simple case of an initially  Gaussian $\zeta$.
Our analysis shows the non-linear signature of Einstein's gravity in large-scale structure.
\end{abstract}

\maketitle

%%==============================%%

%%%%%%%%%%%%%%%%%%%%%%%%%%%%%%%%%%%%%%%%%%%%
\section{Introduction: The gradient expansion in $\Lambda$CDM}
\label{sec:intro}
%%%%%%%%%%%%%%%%%%%%%%%%%%%%%%%%%%%%%%%%%%%%

Einstein's general relativity provides a coherent, causal framework in
which to describe classical cosmological dynamics.
$\Lambda$CDM cosmology is a remarkably successful model for our
observed universe, 
based on a spatially-flat
Friedmann-Lema\^itre-Robertson-Walker (FLRW) spacetime containing 
non-relativistic, collisionless (cold) dark matter (CDM) and a
cosmological constant ($\Lambda$). 
While the homogeneous and isotropic FLRW background can be
studied analytically (using relativistic or Newtonian theory) the
fully non-linear evolution of the 
inhomogeneous matter distribution in
$\Lambda$CDM cosmology 
is usually studied using Newtonian N-body simulations.
These are used to make
detailed predictions for comparison against large-scale galaxy
surveys. As the scale and accuracy of these surveys, and hence that
required from numerical simulations, continues to improve, there has
been growing scrutiny of the reliability of results derived from
Newtonian gravity
\citep{Wands:2009ex,Chisari:2011iq,Green:2011wc,Bruni:2013mua}.  
In this letter we examine the characteristic signature of
general relativity in the large-scale matter density and hence the
galaxy distribution. 

Within $\Lambda$CDM cosmology the distribution of matter is
described by a pressureless fluid.
The relativistic and Newtonian descriptions of the fluid are very
similar if  the appropriate variables are used.   
The fluid is characterised by its density, $\rho$, and its motion, 
represented
by kinematical variables related to its velocity at every point;
$\Theta$ describes the expansion and $\sigma$ its anisotropic deformation or
shear.  
We neglect vorticity to be consistent with the predictions of inflationary cosmology at early times.
If we work in terms of the matter density seen by comoving observers,
$\rho \equiv T_{\mu\nu}u^\mu u^\nu$, and the expansion of the matter
4-velocity, $\Theta\equiv\nabla^\mu u_\mu$, then the relativistic and Newtonian
evolution equations are formally exactly the same \citep{Ell71,maartens:2012}.  
The continuity equation for the matter density is
\beq
\label{covariant:cont}
\dot{\rho} + \Theta \rho  = 0\,,
\eeq
while the Raychaudhuri equation for the expansion is
\beq
\label{covariant:Ray}
\dot{\Theta} + \frac13 \Theta^2 + 2\sigma^2 + 4 \pi G \rho - \Lambda = 0\,,
\eeq
where a dot denotes derivatives with respect to proper time of the comoving observers, corresponding to a Lagrangian time derivative in the Newtonian description.

The difference between Newton and Einstein formalisms becomes evident in
the constraint equations.  
At the heart of Newtonian gravity is the Poisson equation
\beq
\label{poisson}
\nabla^2_{\rm p} \phi = 4\pi G  \rho\;,
\eeq
where $\nabla^2_{\rm p}$ is the spatial Laplacian in physical coordinates.
This gives a {\em linear} relation between the gravitational field, $\phi$, and the matter density, $\rho$. 
In general relativity the density and expansion are related to the intrinsic curvature of the 3-dimensional space orthogonal to $u^\mu$,
denoted by ${}^{(3)}R$. 
This is the energy constraint equation
from Einstein's equations
\beq 
\label{energy:constraint}
\frac23 \Theta^2 - 2\sigma^2 + {}^{(3)}R = 16\pi G \rho + 2 \Lambda\,.
\eeq  

For the homogeneous and isotropic ($\sigma=0$) background, we have $\rho=\bar{\rho}(t)$ and $\Theta=3H(t)$, where $H$ is the Hubble expansion.
The evolution equations \eqref{covariant:cont} and \eqref{covariant:Ray} then become the familiar FLRW equations
\begin{align}
\dot{\bar{\rho}}=&-3H\bar{\rho},\\
\dot{H}=-H^2&-\frac{4\pi G}{3}\bar{\rho} +\frac{\Lambda}{3},
\end{align}

\noindent while the energy constraint \eqref{energy:constraint} reduces to the
Friedmann constraint, 
\beq
\label{friedmann}
H^2=\frac{8\pi G}{3}\bar{\rho} + \frac{\Lambda}{3}\,,
\eeq
with ${}^{(3)}R=0$ for a spatially-flat cosmology. 
We characterise this background model by the present day value of 
the dimensionless density parameter $\Omega_{m}\equiv{8\pi G
  \bar{\rho}_0}/{3 H_0^2}$.

Considering inhomogeneities about the FLRW cosmology, we have
\beqa
\Theta(t,x^i) &=& 3H(t) + \theta (t,x^i)\,, \\
\rho (t,x^i) &=& \bar{\rho}(t) \left[ 1+\delta (t,x^i) \right] \,.
\eeqa
and the inhomogeneous metric can be written in comoving-synchronous coordinates as  
\beq
\label{metric}
ds^2=-dt^2 +a^2(t) e^{2\zeta(t,x^i)}\, \gamma_{kj} dx^k dx^j \,, 
\eeq
where $a(t)$ is the cosmological scale factor and $\gamma_{ij}(t,x^i)$ has unit determinant.

The specific initial conditions for $\Lambda$CDM cosmology
are set by a period of inflation in the very early universe.
In particular inflation produces an almost scale-invariant distribution for the primordial metric perturbation $\zeta$ in Eq.~(\ref{metric})
 \citep{Lyth:2009zz}. This allows us to consider initially small inhomogeneities on large scales, and to perform a gradient expansion (or long-wavelength approximation)
\citep{Lifshitz:1963ps,Tomita:1975kj,Lyth:1984gv,Salopek:1990jq,Deruelle:1994iz,Bruni:2003hm,Rampf:2012pu},
keeping only leading-order terms, i.e., terms at most second order 
in spatial gradients of the metric, in particular $\zeta$.
In this approximation we 
have\footnote{Equation (\ref{gradient:order}) refers to quantities defined in comoving-synchronous coordinates. In particular it is the comoving matter density contrast, $\delta$, that determines the growth of large-scale structure \citep{Wands:2009ex}.
The perturbed expansion $\theta$ and the shear $\sigma$ are the trace and traceless scalars of the deformation tensor, which is equivalent to the extrinsic curvature in our gauge.  This curvature tensor is given by two spatial gradients of the metric and thus both the perturbed expansion and shear are of second order \cite{Tomita:1975kj}.
}
 \citep{Bruni:2013qta}
\beq
\label{gradient:order}
\delta \sim \theta \sim \sigma \sim {}^{(3)}R \sim \nabla^2\,,
\eeq
where $\nabla$ is the spatial gradient in comoving coordinates.
We emphasise that $\delta$, $\theta$ and ${}^{(3)}R$ contain {\em all orders} in a conventional perturbative expansion.
They are leading-order quantities only in terms of spatial gradients.

With this proviso $\delta$ and $\theta$ satisfy the simple
evolution equations, from \eqref{covariant:cont} and
\eqref{covariant:Ray}, 
\beqa
\label{large:continuity}
\dot\delta + \theta &=& {\cal O}(\nabla^4)\,,
 \\
 \label{large:Ray}
\dot\theta +2H\theta +4\pi G \bar\rho\delta &=& {\cal O}(\nabla^4)\,.
\eeqa
These quantities are 
subject to the energy constraint, from \eqref{energy:constraint},
\beq 
\label{large:constraint}
\frac{{}^{(3)}R}{4} +H\theta = 4\pi G \bar\rho \delta + {\cal O}(\nabla^4) \,.
\eeq  
Taking the time derivative of this equation and using the evolution
equations for $\theta$ and $\delta$ we generalise the well-known
first-order result that the conformal curvature,  
\beq
 \label{rescaledR}
R \equiv {}^{(3)}R a^2\,,
\eeq
remains constant in this large-scale limit
\citep{Lukash:1980iv,Lyth:1984gv,Bruni:1992dg}, i.e.\ $R$ is a first
integral of\eqref{large:continuity} and \eqref{large:Ray} .  

%%%%%%%%%%%%%%%%%%%%%%%%%%%%%%%%%%%%
\section{The relativistic growing-mode from the non-linear curvature}
\label{sec:curvature}
%%%%%%%%%%%%%%%%%%%%%%%%%%%%%%%%%%%%

The equations \eqref{large:continuity}, \eqref{large:Ray} and
\eqref{large:constraint} are well known in perturbation theory: they
are the same linear differential equations and constraint that can be
derived at first order in a conventional perturbative expansion in
synchronous-comoving gauge \citep{Bruni:2013qta}, or in a covariant  
gauge-invariant fashion for corresponding quantities
\citep{Bruni:1992dg}. Their solution is therefore  formally the same
than in first-order perturbation theory. The two independent solutions
of these linear differential equations are a decaying and a growing
mode \citep{Pee80}. Thanks to inflation, the decaying mode is
negligible, so we focus on the the growing-mode solution.  In the
large-scale limit we thus have \citep{Bruni:2013qta} 
\beq
 \label{large:delta}
\delta = C(x^i) D_+(t) \,, \quad  
\theta = -C(x^i) \dot{D}_+(t),
\eeq
where the growth factor, $D_+(t)$, is proportional to the
scale factor, $a(t)$, in an Einstein-de Sitter ($\Omega_m=1$)
cosmology \citep{Bernardeau:2001qr}.  
The growing-mode amplitude, $C(x^i)$, is related to the conformal
curvature on large scales through the energy constraint equation
\eqref{large:constraint} evaluated at an initial time $t_{IN}$ early in
the matter-dominated era, 
\beq
 \label{large:amplitude}
C(x) = \frac{R}{10a_{IN}^2H_{IN}^2 D_{+IN}} \,.
\eeq

The growing mode solution  for $\delta$ and
$\theta$ on large scales, Eq.\ \eqref{large:delta}, has the same time dependence as the
first-order perturbative solution, thus it may referred to as the linearly growing mode, however we remark again that  in our non-linear case $R$ is only conserved at leading order in our gradient expansion\footnote{In a conventional perturbative expansion, for pressureless matter
$R$ is conserved at all scales at first order \citep{Bruni:1992dg}, but at second order contains a time-dependent part that can be neglected at large scales  \citep{Bruni:2013qta}.}; 
$\delta$,
$\theta$ $R$ and $C$ here contain the large-scale part of all perturbative orders. 

In single-field, slow-roll inflation, the primordial metric
perturbation, $\zeta(t,x^i)$, is predicted to have an almost Gaussian
distribution \citep{Maldacena:2002vr,Acquaviva:2002ud}. 
Crucially, the conformal curvature $R$, Eq.\ \eqref{rescaledR}, is a {\em non-linear}
function of the spatial metric 
 in Eq.~\eqref{metric}.
Considering only the scalar part of the initial metric
perturbation at leading order on large scales then the spatial metric
 can be taken to have a simplified form\footnote{For scalar perturbations the non-Euclidean part of $\gamma_{ij}$ would be of order $\nabla^2$, hence these terms would give contributions to $R$ of order $\nabla^4$.},
$\gamma_{ij}\simeq\delta_{ij}$, and the conformal curvature
$R$ is then a non-linear function of only the 
perturbation $\zeta$ in \eref{metric}. With $\gamma_{ij}\simeq\delta_{ij}$ the function $a(t)\exp[\zeta(x)]$ effectively acts as a local scale factor in the so called ``separate universe" picture corresponding to our gradient expansion. $R$ then represents the corresponding local spatial curvature and 
 takes a beautifully simple and exact form 
\citep{Wald:1984rg}    
\beqa
\label{conformal:ricci}
R &\simeq&  \exp{(-2\zeta)} \left[ -4 \nabla^2 \zeta - 2 \left( \nabla
  \zeta \right)^2 \right] \,
\nonumber
\\
&\simeq & - 4 \Big[ \nabla^2\zeta + \frac12 \left( \nabla \zeta \right)^2
  -2 \zeta \nabla^2 \zeta - \nonumber \\
&&\,\zeta \left( \nabla \zeta \right)^2 + 2
  \zeta^2 \nabla^2 \zeta +\ldots \Big] \,. 
\eeqa
This expression is second-order in spatial gradients, consistent with
\eqref{gradient:order}, but non-linear in terms of the metric perturbation, 
$\zeta$.
Consequently, even if $\zeta$ is described by a Gaussian distribution,
its non-linear relation to the curvature $R$ leads to a non-Gaussian
distribution \citep{inprep} for the comoving density contrast, $\delta$, determined
by the amplitude \eqref{large:amplitude} of the growing mode
\eqref{large:delta}.  

At first order in a perturbative expansion 
we have
from \eref{conformal:ricci} 
\beq
\label{first:R}
R_1 = -4\nabla^2 \zeta_1 \,.
\eeq
Substituting this into the constraint equation (\ref{large:constraint}) we recover the
Poisson equation (\ref{poisson}),
where we identify the Newtonian potential in terms of the first-order
Ricci curvature and the inhomogeneous expansion in the
comoving-synchronous gauge  
\beqa
 \nabla^{2} \phi_1 &=& a^2 \left[ \frac14 {}^{(3)}R_1 + H\theta_1 \right] 
\nonumber\\
&=& 
 - \nabla^{2} \zeta_1 + a^2 H \theta_1 \,.
\eeqa

Using the full non-linear expression in \eref{conformal:ricci}, we
can write the conformal curvature in terms of $\zeta$ as an infinite series
\beqa
&&R \simeq  - 4 \nabla^2\zeta + \nonumber \\
&&\quad\sum_{m = 0}^{\infty}
\frac{(-2)^{m+1}}{m!}  \left[ (m+1) \left( \nabla \zeta \right)^2 -4 \zeta \nabla^2 \zeta \right] \zeta^m .
\eeqa
It is this non-linear curvature which determines the non-linear
amplitude (\ref{large:amplitude}) of the growing mode density
perturbation (\ref{large:delta}). 

%%%%%%%%%%%%%%%%%%%%%%%%%%%%%%%%%%%%%%%%%%%%
% 
\section{The non--linear relativistic effect on structure formation}
\label{sec:deltanl}
%%%%%%%%%%%%%%%%%%%%%%%%%%%%%%%%%%%%%%%%%%%%

We wish to relate this density contrast to the observed
distribution of galaxies revealed by astronomical surveys. 
Although a full description requires complex, non-linear astrophysics we can assume that 
in $\Lambda$CDM cosmology, galaxies form in virialised dark matter halos which are
biased tracers of the underlying matter distribution on large scales \citep{Peacock:1999ye}.  
In the simplest model of spherical collapse in Einstein-de Sitter,
written in comoving-synchronous coordinates, there is an exact
parametric solution \citep{Pee80}, 
\beqa
 \delta & = & \frac{9(\psi-\sin\psi)^2}{2(1-\cos\psi)^3} - 1 \,, \\
 t & = & \frac{6^{3/2}}{2R^{3/2}} \left( \psi - \sin\psi \right) \,,
 \eeqa
which can be expanded term by term as 
\beq
 \delta = C D_+ + \frac{38}{21} (C D_+)^2 + \ldots \,.
\eeq
where the linearly growing mode \eqref{large:delta}, with \eqref{large:amplitude},  is given
by $CD_+ = Ra/10$ for Einstein-de Sitter 
in both Newtonian theory and
general relativity \citep{Wands:2009ex}. Halos collapse when $\psi=2\pi$ and the linearly
evolved density contrast reaches a critical value
$\delta_*=1.686$. Thus we can predict the number of collapsed halos 
(of a given mass) 
at a given time in terms of the number of peaks of the
initial growing mode of the comoving density contrast 
(smoothed on a given mass scale) 
above a critical value \citep{Press:1973iz}.
Going beyond the
spherical collapse, this is the barrier crossing approach, where halos form
where the linearly growing mode exceeds a critical value.
 
Note that it is the {\em non-linear} amplitude, $C$, of the 
{\em linearly evolved} growing mode \eqref{large:delta} which
determines the halo density and this is given by the full non-linear
conformal curvature, $R$ in \eref{conformal:ricci}. In a general
relativistic description of spherical collapse \citep{Wands:2009ex} it
is thus the initial density contrast in the local comoving matter,
$\delta$, that predicts the distribution of halos \citep{Bruni:2011ta}
and as we have seen, this is non-linearly related to the primordial
metric perturbation $\zeta$.

To understand the effect of this non-linearity on structure formation,
we consider a peak-background split \citep{Peacock:1999ye}, where one
decomposes a field into shorter wavelength modes, which generate local
peaks, and much longer 
wavelength modes which modulate the number density of peaks\footnote
{Equivalent conclusions can be obtained by studying the distribution of peaks of the metric perturbation, setting $(\nabla\zeta)^2=0$, or by studying the squeezed limits of higher-order correlation functions of the density field \citep{Bruni:2013qta}.
}.
Note that, since we have already made a gradient expansion in the
above (wavenumbers $k<k_{\rm max}$), spatial gradients of our ``shorter wavelength'' modes should
still be small ($k_{\rm split}<k<k_{\rm max}$), and we will now drop completely all gradients of the very
long wavelength modes ($k<k_{\rm split}$).  

 For simplicity, from now on we shall assume the simplest inflationary scenario where $\zeta$ is Gaussian, focusing on the specific general-relativistic non-Gaussianity introduced by the  non-linearity of \eref{conformal:ricci} in the constraint \eqref{large:constraint}.
 We can split
\beq
\label{split}
\zeta \equiv \zeta_\ell + \zeta_s \,,
\eeq
where the longer and shorter wavelength modes are
independent for an
initially Gaussian metric perturbation. Substituting \eref{split} into \eref{conformal:ricci} 
we obtain 
\beqa
&&R \simeq \exp(-2\zeta_\ell) R_s + \nonumber \\
&&\quad 4 \exp(-2\zeta_\ell-2\zeta_s) \nabla\zeta_\ell\nabla\zeta_s +  \exp(-2\zeta_s) R_\ell ,
\eeqa
where
\beq
 \label{short:ricci}
 R_s = \exp{(-2\zeta_s)} \left[ -4 \nabla^2 \zeta_s - 2 \left( \nabla \zeta_s \right)^2 \right] \,,
 \eeq
and similarly for $R_\ell$.
Dropping all spatial gradients of long-wavelength modes,
$\nabla\zeta_\ell$, i.e., taking these modes to define a locally
homogeneous background, we find that the spatial curvature due to
short wavelength modes is modulated such that $R\simeq
\exp(-2\zeta_\ell)R_s$. 
This is consistent with the interpretation that the long wavelength
metric perturbation is a rescaling of the local
background scale factor \citep{Maldacena:2002vr,Creminelli:2004yq,Bartolo:2005fp,Creminelli:2011sq,Creminelli:2013mca}
\beq
 \label{local:scale}
a \to a_\ell = \exp(\zeta_\ell) a
 \,.
\eeq
Hence the local amplitude of the 
growing mode of the density
contrast is also modulated (cf. \eqref{large:delta} and
\eqref{large:amplitude})  
\beq
 \label{short:delta}
 \delta  = \exp(-2\zeta_\ell) \delta_s  +  \mathcal{O}(\nabla\zeta_\ell)\,.
\eeq
The non-linear effect of a long-wavelength overdensity,
$\zeta_\ell >0$, suppresses the amplitude of
shorter-wavelength modes since $\zeta_\ell>0$ increases the local
effective scale factor, suppressing spatial curvature and thus the
density contrast.  

We can compare \eref{short:delta} with local-type primordial
non-Gaussianity \citep{Wands:2010af} in a Newtonian approach where the
amplitude of the linearly growing mode of the density is determined by
the Newtonian potential 
\beq
 \label{local:phi}
\phi = \phi_1 + f_{\rm NL} \left( \phi_1^2 - \langle \phi_1^2 \rangle \right) 
 + g_{\rm NL} \phi_1^3 + h_{\rm NL} \left( \phi_1^4 - \langle \phi_1^4 \rangle \right) + \ldots \,.
\eeq
If we split the first-order Newtonian potential into longer and shorter
wavelength modes $\phi_1=\phi_\ell+\phi_s$ and drop the gradients of
$\phi_\ell$, we find 
\beq
 \label{local:delta}
\delta = \left( 1 + 2f_{\rm NL} \phi_\ell + 3g_{\rm NL} \phi_\ell^2 
+ 4h_{\rm NL} \phi_\ell^3 +\ldots \right) \delta_s  + \ldots
\eeq
The modulation of the amplitude of smaller-scale density fluctuations
$\delta_s$ by the long-wavelength potential, $\phi_\ell$, modifies the
halo density giving rise to a strong modulation of the halo power
spectrum on sufficiently large scales, where $\phi_\ell$ remains
finite even though the long-wavelength density contrast is suppressed,
$\delta\sim\nabla^2$. This leads to a scale-dependent bias in the
distribution of galaxies on large scales 
\citep{Dalal:2007cu,Matarrese:2008nc}.  

If we impose the same linear relation between $\zeta$ and the
Newtonian potential, $\phi=(3/5)\zeta$,
 that is valid for first-order perturbations in the matter-dominated era, 
then in single-field, slow-roll inflation $f_{\rm NL}$ and all
higher-order coefficients are suppressed. This results in an
effectively Gaussian distribution for the Newtonian potential and 
hence (in Newtonian theory) the density field. However, expanding the
exponential in \eref{short:delta} and comparing term by term with the
equivalent Newtonian expression \eref{local:delta} we can identify 
the effective non-Gaussianity on large scales in
general relativity 
\beq
 \label{eff:fNL}
f_{\rm NL}^{\rm GR}= -\frac53
  \,,\quad
 g_{\rm NL}^{\rm GR}= \frac{50}{27}
 \,, \quad
 h_{\rm NL}^{\rm GR}= -\frac{125}{81}
 \,, \cdots
 \eeq
More generally we find
\beq \label{gen:fNL}
f_{\rm NL}^{(n)\ {\rm GR}}= \frac{1}{n!} \left( -\frac{10}{3} \right)^{n-1} \,,
\eeq 
where we write the local expansion (\ref{local:phi}) as
\beq\label{gen:phiNL}
 \phi = \phi_1 + \sum_{n=2}^\infty f_{\rm NL}^{(n)} \left( \phi_1^n - \langle\phi_1^n \rangle \right) \,,
\eeq
extending the previous result at second order for $f_{\rm NL}^{\rm
  GR}$ \citep{Bartolo:2005xa,Verde:2009hy,Bartolo:2010rw,Hidalgo:2013,Bruni:2013qta,Uggla:2014hva} to higher
orders. 

%%%%%%%%%%%%%%%%%%%%%%%%%%%%%%%%%%%%%%%%%%%%
\section{Discussion}
\label{sec:disc}
%%%%%%%%%%%%%%%%%%%%%%%%%%%%%%%%%%%%%%%%%%%%

% 
Traditionally, primordial non-Gaussianity is described in terms of the Newtonian gravitational potential, $\phi$ [for example equation (\ref{gen:phiNL})], linearly related to the density field through the Poisson equation (\ref{poisson}). On the other hand, inflationary predictions are expressed in terms of the primordial metric perturbation, $\zeta$ in Eq.~(\ref{metric}).
Our results, valid in full non-linearity and at large scales,  
show how the essential non-linearity of general relativity produces an intrinsic non-Gaussianity in the matter density field and hence the galaxy distribution on large scales, even starting from purely Gaussian primordial metric perturbations, generalising previous results in second-order perturbation theory \citep{Tomita:2005et,Bartolo:2005xa,Verde:2009hy,Bartolo:2010rw,Hwang:2012bi,Hidalgo:2013,Bruni:2013qta,Uggla:2014hva}
We also need a detailed modelling of observational surveys, including all geometrical and relativistic effects, to fully disentangle effects of primordial non-Gaussianity from intrinsic non-linearity in general relativity \citep{Bruni:2011ta,Raccanelli:2013dza,inprep}. 
Most studies of GR effects on observations of large-scale structure have been restricted to linear perturbation theory \citep{Yoo:2009au,Yoo:2010ni,Bonvin:2011bg,Challinor:2011bk},  
but there have been recent attempts to include non-linear effects, see e.g.\  \citep{Thomas:2014aga,Bertacca:2014dra,Bertacca:2014wga,Yoo:2014sfa,DiDio:2014lka,Jeong:2014ufa}.

Alternative gravity theories may impose different constraints between
the primordial metric perturbation $\zeta$ and the comoving density
contrast $\delta$, and hence could in principle be distinguished by a
different galaxy distribution on large scales.
This could be an interesting approach to testing gravity on
cosmological scales,  
complementary to existing work which probes gravity through the growth
of cosmic structure at late times \citep{Zhao:2011te}. 

Even within the context of general relativity the constraint equation
(\ref{energy:constraint}) could include additional contributions due
to other fields such as dark energy/quintessence. 
Fields which have a negligible effect in the background could still
contribute to the inhomogeneous perturbations, e.g., magnetic fields
or gravitational waves. In particular we have considered only the
growing mode of scalar perturbations at early times. Tensor metric
perturbations are decoupled from scalar density perturbations at first
order, but do contribute to the Ricci curvature at second order, even
in the large scale limit, and hence could contribute to the non-linear
density perturbation \citep{Matarrese:1997ay,Dai:2013kra}, although this is expected to be sub-dominant.  

In summary, in this letter we have obtained for the first time the fully non-linear general relativistic 
initial distribution of primordial density perturbations in $\Lambda$CDM 
on large scales 
\beq
\delta= \frac{  \exp{(-2\zeta)} \left[ -4 \nabla^2 \zeta - 2 \left( \nabla
  \zeta \right)^2 \right]  }{10a_{IN}^2H_{IN}^2 D_{+IN}} D_+(t)\,,
\eeq
an expression including {\em  all perturbative orders}. 
This fully non-linear relation between $\delta$ and $\zeta$ clearly shows that, even 
for a Gaussian-distributed $\zeta$, 
the corresponding matter density field is non-Gaussian.
Assuming simple inflationary Gaussian initial conditions in $\zeta$,
and using a peak-background split,  we have derived the corresponding specific general-relativistic effective non-Gaussianity 
parameters, \eref{eff:fNL} and \eref{gen:fNL}, that results when a Newtonian treatment is used, i.e. a Poisson equation as relation between $\delta$ and the Newtonian potential $\phi$, and a linear relation is assumed between $\zeta$ and $\phi$.
Although Newtonian simulations are commonly used to study the
non-linear evolution of the matter density contrast $\delta$, a
relativistic approach is essential to properly determine the initial
conditions set
by a period of inflation in the very early
universe. Thus we have shown how Einstein's gravity imprints a characteristic signature
in the large-scale structure of our universe.  
\newpage
%%%%%%%%%%%%%%%%%%%%%%%%%%%%%%%%%%%%%%%%%%%%
\section{Acknowledgements}
\label{sec:thanks}
%%%%%%%%%%%%%%%%%%%%%%%%%%%%%%%%%%%%%%%%%%%%

\noindent The authors are grateful to Rob Crittenden, Roy Maartens and Gianmassimo Tasinato for useful 
{discussions}. 
This work was supported by STFC grants ST/K00090X/1 and ST/L005573/1, and by PAPIIT-UNAM grants IN103413-3 and IA101414-1.

%%%==================================%%%

%%%==================================%%%


\begin{thebibliography}{}
\expandafter\ifx\csname natexlab\endcsname\relax\def\natexlab#1{#1}\fi
%12
\bibitem[{Bruni {et~al.}(2014)Bruni, Thomas, \& Wands}]{Bruni:2013mua}
Bruni, M., Thomas, D.~B., \& Wands, D. 2014, Phys.Rev., D89,044010,  arXiv:1306.1562 

%14
\bibitem[{Chisari \& Zaldarriaga(2011)}]{Chisari:2011iq}
Chisari, N.~E., \& Zaldarriaga, M. 2011, Phys.Rev., D83, 123505,  arXiv:1101.3555

%23
\bibitem[{Green \& Wald(2012)}]{Green:2011wc}
Green, S.~R., \& Wald, R.~M. 2012, Phys.Rev., D85, 063512, arXiv:1111.2997

%46
\bibitem[{Wands \& Slosar(2009)}]{Wands:2009ex}
Wands, D., \& Slosar, A. 2009, Phys.Rev., D79, 123507,  arXiv:0902.1084

%22
\bibitem[{{Ellis}(1971)}]{Ell71}
{Ellis}, G.~F.~R. 1971, in General Relativity and Cosmology, ed. {R.~K.~Sachs},
  Proc. Int. School of Physics `Enrico Fermi' (Varenna), Course XLVII, (New
  York: Academic Press), 104--182. {\it Republished in:} Gen.Rel.Grav. {\bf 41}
  581--660 (2009)

%22bis
\bibitem[{{Ellis} {et~al.}(2012){Ellis}, {Maartens}, \& H.}]{maartens:2012}
{Ellis}, G.~F.~R., {Maartens}, R., \& {MacCallum}, M.~A.~H. 2012, {Relativistic
  Cosmology} (Cambridge University Press)

%30
\bibitem[{{Lyth} \& {Liddle}(2009)}]{Lyth:2009zz}
{Lyth}, D.~H., \& {Liddle}, A.~R. 2009, {The Primordial Density Perturbation}
  (Cambridge University Press)
 
%11
\bibitem[Bruni \& Sopuerta(2003)]{Bruni:2003hm} Bruni, M., \& Sopuerta, C.~F.\ 2003, Classical and Quantum Gravity, 20, 5275, arXiv: gr-qc/0307059

%20
\bibitem[Deruelle 
\& Langlois(1995)]{Deruelle:1994iz} Deruelle, N., \& Langlois, D.\ 1995, \prd, 52, 2007, arXiv: gr-qc/9411040

%27
\bibitem[{{Lifshitz}\& {Khalatnikov} (1963)}]{Lifshitz:1963ps} 
  {Lifshitz},E.~M., \& {Khalatnikov}I.~M. 1963,
  {Investigations in relativistic cosmology,}
  Adv.\ Phys.\  {\bf 12}, 185

%29
\bibitem[{Lyth(1985)}]{Lyth:1984gv}
Lyth, D. 1985, Phys.Rev., D31, 1792

%38
\bibitem[{Rampf \& Rigopoulos(2013)}]{Rampf:2012pu}
Rampf, C., \& Rigopoulos, G. 2013, Mon.Not.Roy.Astron.Proc., 430, L54, arXiv:1210.5446

%39
\bibitem[Salopek 
\& Bond(1990)]{Salopek:1990jq} Salopek, D.~S., \& Bond, J.~R.\ 1990, 
\prd, 42, 3936   

 %41
\bibitem[{{Tomita}(1975)}]{Tomita:1975kj}
{Tomita}, K. 1975, Progress of Theoretical Physics, 54, 730

%10
\bibitem[Bruni et al.(2014)]{Bruni:2013qta} Bruni, M., Hidalgo, 
J.~C., Meures, N., \& Wands, D.\ 2014, \apj, 785, 2,  arXiv:1307.1478

%8
\bibitem[Bruni et al.(1992)]{Bruni:1992dg} Bruni, M., Dunsby, 
P.~K.~S., \& Ellis, G.~F.~R.\ 1992, \apj, 395, 34, 

 %28
\bibitem[Lukash(1980)]{Lukash:1980iv} Lukash, V.~N.\ 1980, Soviet 
Journal of Experimental and Theoretical Physics, 52, 807 

%35
\bibitem[{{Peebles}(1980)}]{Pee80}
{Peebles}, P.~J.~E. 1980, {The large-scale structure of the universe}
  (Princeton University Press)

%5
\bibitem[{Bernardeau {et~al.}(2002)Bernardeau, Colombi, Gaztanaga, \&
  Scoccimarro}]{Bernardeau:2001qr}
Bernardeau, F., Colombi, S., Gaztanaga, E., \& Scoccimarro, R. 2002,
  Phys.Rept., 367, 1, arXiv: astro-ph/0112551

%1
\bibitem[Acquaviva et al.(2003)]{Acquaviva:2002ud} Acquaviva, V., 
Bartolo, N., Matarrese, S., 
\& Riotto, A.\ 2003, Nuclear Physics B, 667, 119, arXiv: astro-ph/0209156

%31
\bibitem[{Maldacena(2003)}]{Maldacena:2002vr}
Maldacena, J.~M. 2003, JHEP, 0305, 013, arXiv: astro-ph/0210603

%45
\bibitem[{{Wald}(1984)}]{Wald:1984rg}
{Wald}, R.~M. 1984, {General relativity} (University of Chicago Press)

%26
 \bibitem[{Koyama} (2014, in preparation)]{inprep}
 Koyama, K., Maartens, R., and Wands, D. (2014) {\em in preparation}.

%34
\bibitem[Peacock(1999)]{Peacock:1999ye} Peacock, J.~A.\ 1999, 
Cosmological Physics, by John A.~Peacock, pp.~704.~ISBN 
052141072X.~Cambridge, UK: Cambridge University Press, January 1999., 

%36
\bibitem[Press 
\& Schechter(1974)]{Press:1973iz} Press, W.~H., \& Schechter, P.\ 1974, \apj, 187, 425 

%9
\bibitem[{{Bruni} {et~al.}(2012){Bruni}, {Crittenden}, {Koyama}, {Maartens},
  {Pitrou}, \& {Wands}}]{Bruni:2011ta} 
{Bruni}, M., {Crittenden}, R., {Koyama}, K., {et~al.} 2012, \prd, 85, 041301, arXiv:1106.3999

%4
\bibitem[Bartolo et al.(2005)]{Bartolo:2005fp} Bartolo, N., Matarrese, 
S., \& Riotto, A.\ 2005, \jcap, 8, 10, arXiv: astro-ph/0506410

%15
\bibitem[Creminelli et al.(2013)]{Creminelli:2013mca} Creminelli, P., 
Nore{\~n}a, J., Simonovi{\'c}, M., \& Vernizzi, F.\ 2013, \jcap, 12, 25, arXiv:1309.3557

%16
\bibitem[Creminelli et al.(2011)]{Creminelli:2011sq} Creminelli, P., 
Pitrou, C., \& Vernizzi, F.\ 2011, \jcap, 11, 25, arXiv:1109.1822

%17
\bibitem[Creminelli 
\& Zaldarriaga(2004)]{Creminelli:2004yq} Creminelli, P., \& Zaldarriaga, M.\ 2004, \jcap, 10, 6,
arXiv: astro-ph/0407059

%47
\bibitem[{Wands(2010)}]{Wands:2010af}
Wands, D. 2010, Class.Quant.Grav., 27, 124002, arXiv:1004.0818

%19
\bibitem[{Dalal {et~al.}(2008)Dalal, Dore, Huterer, \& Shirokov}]{Dalal:2007cu}
Dalal, N., Dore, O., Huterer, D., \& Shirokov, A. 2008, Phys.Rev., D77, 123514, arXiv:0710.4560 

%33
\bibitem[{Matarrese \& Verde(2008)}]{Matarrese:2008nc}
Matarrese, S., \& Verde, L. 2008, Astrophys.J., 677, L77, arXiv:0801.4826

%2
\bibitem[{{Bartolo} {et~al.}(2010){Bartolo}, {Matarrese}, {Pantano}, \&
  {Riotto}}]{Bartolo:2010rw}
{Bartolo}, N., {Matarrese}, S., {Pantano}, O., \& {Riotto}, A. 2010, Classical
 and Quantum Gravity, 27, 124009, arXiv:1002.3759 

%2bis
\bibitem[Hwang et al.(2012)]{Hwang:2012bi} Hwang, J.-c., Noh, H., 
\& Gong, J.-O.\ 2012, \apj, 752, 50, arXiv:1204.3345  
 
%3
\bibitem[{Bartolo {et~al.}(2005)Bartolo, Matarrese, \& Riotto}]{Bartolo:2005xa}
Bartolo, N., Matarrese, S., \& Riotto, A. 2005, JCAP, 0510, 010, arXiv: astro-ph/0501614

%24
\bibitem[Hidalgo et al.(2013)]{Hidalgo:2013} Hidalgo, J.~C., 
Christopherson, A.~J., \& Malik, K.~A.\ 2013, \jcap, 8, 26, arXiv:1303.3074

%43
\bibitem[{Verde \& Matarrese(2009)}]{Verde:2009hy}
Verde, L., \& Matarrese, S. 2009, Astrophys.J., 706, L91, arXiv:0909.3224

%44
\bibitem[Uggla 
\& Wainwright(2014)]{Uggla:2014hva} Uggla, C., \& Wainwright, J.\ 2014, arXiv:1402.2464 

%42
\bibitem[Tomita(2005)]{Tomita:2005et} Tomita, K.\ 2005, \prd, 71, 
083504, arXiv: astro-ph/0501663

%37
\bibitem[Raccanelli et al.(2013)]{Raccanelli:2013dza} Raccanelli, A., 
Bertacca, D., Dore, O., \& Maartens, R.\ 2014,  \jcap, 8, 22 arXiv:1306.6646 

%7
\bibitem[Bonvin 
\& Durrer(2011)]{Bonvin:2011bg} Bonvin, C., \& Durrer, R.\ 2011, \prd, 84, 063505,
arXiv:1105.5280

%13
\bibitem[Challinor 
\& Lewis(2011)]{Challinor:2011bk} Challinor, A., \& Lewis, A.\ 2011, \prd, 84, 043516, arXiv:1105.5292

%48
\bibitem[Yoo et al.(2009)]{Yoo:2009au} Yoo, J., Fitzpatrick, 
A.~L., \& Zaldarriaga, M.\ 2009, \prd, 80, 083514, arXiv:0907.0707

%48bis
\bibitem[Yoo(2010)]{Yoo:2010ni} Yoo, J.\ 2010, \prd, 82, 083508, arXiv:1009.3021

%6
\bibitem[Bertacca et al.(2014a)]{Bertacca:2014dra} Bertacca, D., 
Maartens, R., \& Clarkson, C.\ 2014, \jcap, 09, 37, arXiv:1405.4403 

%6bis
\bibitem[Bertacca et al.(2014b)]{Bertacca:2014wga} Bertacca, D., 
Maartens, R., \& Clarkson, C.\ 2014, arXiv:1406.0319 

%21
\bibitem[Di Dio et al.(2014)]{DiDio:2014lka} Di Dio, E., Durrer, R., 
Marozzi, G., \& Montanari, F.\ 2014, arXiv:1407.0376 

%25
\bibitem[Jeong 
\& Schmidt(2014)]{Jeong:2014ufa} Jeong, D., \& Schmidt, F.\ 2014, arXiv:1407.7979 

 %40
\bibitem[Thomas et al.(2014)]{Thomas:2014aga} Thomas, D.~B., Bruni, 
M., \& Wands, D.\ 2014, arXiv:1403.4947 	

%49
\bibitem[Yoo 
\& Zaldarriaga(2014)]{Yoo:2014sfa} Yoo, J., \& Zaldarriaga, M.\ 2014, \prd, 90, 023513, arXiv:1406.4140


%50
\bibitem[Zhao et al.(2012)]{Zhao:2011te} Zhao, G.-B., Li, H., 
Linder, E.~V., et al.\ 2012, \prd, 85, 123546, arXiv:1109.1846

%18
\bibitem[Dai et al.(2013)]{Dai:2013kra} Dai, L., Jeong, D., 
\& Kamionkowski, M.\ 2013, \prd, 88, 043507, arXiv:1306.3985

%32
\bibitem[{{Matarrese} {et~al.}(1998){Matarrese}, {Mollerach}, \&
  {Bruni}}]{Matarrese:1997ay}
{Matarrese}, S., {Mollerach}, S., \& {Bruni}, M. 1998, \prd, 58, 043504, arXiv: astro-ph/9707278





%%%%%%%%%%%%%%%%%%%%%%%%%%% OLD
  
\end{thebibliography}
\end{document}